\documentclass[doublecol]{epl2} 

\usepackage{graphicx}
\usepackage{amsmath}
\usepackage{amssymb}
\usepackage{bm}
\usepackage{amsfonts}
\usepackage{appendix}

\title{Correspondence between many-particle excitations and the entanglement spectrum of disordered ballistic one-dimensional systems}
\shorttitle{Correspondance between excitations and the entaglement spectrum}

\author{Shaul Leiman\inst{1} \and Ariel Eisenbach\inst{1} 
\and Richard Berkovits\inst{1}}
\shortauthor{S. Leiman \etal}

\institute{                    
  \inst{1} Department of Physics,  Jack and Pearl Resnick Institute,
Bar-Ilan University, Ramat-Gan 52900, Israel
}
\pacs{61.43.-j}{Disordered Solids}
\pacs{73.21.Hb}{Quantum Wires}
\pacs{03.65.Ud}{Entanglement and Quantum Non-locality}

\abstract{
Using exact diagonalization for non-interacting systems and density matrix
renormalization group for interacting systems we show that Li and Haldane's 
conjecture on the correspondence between the low-lying many-particle 
excitation spectrum and the entanglement spectrum holds for
disordered ballistic one-dimensional many-particle systems. 
In order to demonstrate
the correspondence we develope a computational
efficient way to calculate the ES of low-excitation of non-interacting
systems. We observe and explain the presence of an unexpected shell structure
in the excitation structure. The low-lying shell are robust and survive even
for strong electron-electron interactions.}

\begin{document}

\maketitle

\section{Introduction}

Concepts developed in the field of quantum information are gaining a foothold
in condensed mater physics \cite{amico08,eisert09}.
One of the most influential concepts is related to 
quantifying the entanglement between two parts of a system.
Specifically, a many particle system in a pure state, may be divided
into two regions A and B.
The entanglement between the regions A and B can be quantified by different
measures such as the entanglement entropy, R\'enyi entropy and
entanglement spectrum (ES),
connected to the reduced density matrix of area A, $\rho_A$ or B, $\rho_B$.

The canonical method for 
utilizing the information embedded in 
the eigenvalues $\lambda_i^{A}$ of $\rho_A$
is the entanglement entropy:
$S_{A}=- \sum_i \lambda_i^{A} \ln \lambda_i^{A}$.
The ES is constructed out of the set of eigenvalues 
$\{\lambda_i^{A}\}$ by the transformation:
$\{\varepsilon_i^A=- \ln \lambda_i^{A}\}$.
Several years ago Li and Haldane \cite{haldane08} came
up with an intriguing conjecture regarding the ES
of a many-body state.
They show that the ES of a
partitioned fractional quantum Hall 
$\nu=5/2$ state resembled the minimal edge
excitation spectrum, and suggest a connection
between the properties of the ES and the topological
order of this state. 

This idea led to many additional studies
\cite{calabrese08,r2,r3,r5,r7,r8,r9,r10,r11,r13,r16,r17,r18,r19,r21,r22,r23,sondhi14}, mainly focused on systems with topological behavior, 
suggesting that the low-energy ES distribution
shows some correspondence to the true many-body
excitations (MBE) of the partitioned segment (region A). 
Since the reduced density matrix of a region is adiabatically
connected to the MBE of the disconnected region, one may expect that 
the exact ground state eigenfunction of the whole system
encodes information on the sub-system's
low lying excitations.

In this letter we would use Li and Haldanes' conjecture
in order to investigate properties of the excitation spectrum
of fermionic disordered
systems. The effect of disorder on the ES of edge states 
in topological
insulators has been previously studied \cite{prodan10,gilbert12}.
Properties of several low lying ES levels have been used to
identify a metallic phase in one-dimensional Bose-Hubbard models
\cite{romer15}.
Nevertheless, a direct comparison between the properties of many particle
excitations in
a weakly disordered systems and the ES of a corresponding segment
is still lacking.

The field of MBE statistics
has a long history, and continues to draw interest
in diverse areas such as
cold atoms in the presence of quasi-periodic potentials 
\cite{roati08,lahini09,tanzi13,errico14},
and the many body localization (MBL) transition 
\cite{basko06,basko07,gornyi05}.
For disordered systems 
it is natural to seek knowledge on the statistical properties of 
spectrum. Specifically, we are interested in answering the question:
Does the ES exhibit statistical properties corresponding to the MBE 
spectrum. 
For the disordered single particle spectrum there is an extensive literature
on the statistical properties of the energy spectrum in the
localized, critical, diffusive and chaotic regimes. 
Different energy spectrum and wave function statistics, depending
on whether the disordered system has time reversal symmetry (Gaussian
orthogonal ensemble (GOE)), broken time reversal symmetry 
(Gaussian unitary statistics (GUE)), spin-orbit
interactions (Gaussian symplectic statistics (GSE)) 
\cite{dyson63, gorkov65,altshuler86}. The energy spectrum statistics can
be used to identify the Anderson localization transition
\cite{shklovskii93}.

The statistics of MBE in disordered interacting 
systems have an interesting twist.
For non-interacting many-particle systems the
level spacing MBE distribution is expected to follow
the Poisson distribution for excitation energies above the second spacing,
without depending on the single-level distribution
\cite{berkovits94}.
On the other hand, once repulsive interactions between the
particles are considered, a transition to the Wigner distribution
for higher excitations is observed
\cite{berkovits96,pascaud98,berkovits99,song00,oganesyan07}.
The main difficulty
in studying this transition 
is that exact diagonalization needed to study excited
states is limited to very small systems. 
Here, the Li and Haldanes' conjecture can come to the rescue, since as we shall
demonstrate below, one may extract the ES up to a few hundreds of states.
Thus, if indeed there is a connection between the ES and The MBE,
the low level excitations of a rather large many-body 
systems are numerically available.

In this letter we shall explicitly demonstrate the correspondence
between the excitation spectrum and the ES, for non-interacting one-dimensional
disordered systems in the chaotic regime. In order to facilitate the calculation
of the ES for large systems, we present a numerical method based on the
correlation matrix eigenvalues, which could be used for non interacting systems
in any dimension. We show that the average entanglement spectrum
level spacings (ESLS) 
show a shell structure, with a large average spacing appearing according to the
combinatorical partition function $p(m)$. The 
ESLS follows the Poisson statistics for the small spacings,
while it shows a narrow distribution for the large spacings. Adding interactions
removes the shell structure for strong interactions at higher levels, but the
shell structure is rather robust for the low-lying portion of the ES, 
corresponding to the MBE which are close to the
Fermi energy. 



\section{Model}
As an example of a disordered many-particle system,
we consider a spinless 1D electrons system of size $L$ 
with repulsive nearest-neighbor
interactions and on-site disordered potential.
The Hamiltonian is given by:
\begin{eqnarray} \label{hamiltonian}
H &=& 
\displaystyle \sum_{j=1}^{L} \epsilon_j {\hat c}^{\dagger}_{j}{\hat c}_{j}
-t \displaystyle \sum_{j=1}^{L-1}({\hat c}^{\dagger}_{j}{\hat c}_{j+1} + h.c.) \\ \nonumber
&+& U \displaystyle \sum_{j=1}^{L-1}({\hat c}^{\dagger}_{j}{\hat c}_{j} - \frac{1}{2})
({\hat c}^{\dagger}_{j+1}{\hat c}_{j+1} - \frac{1}{2}),
\end{eqnarray}
where $\epsilon_j$ is the on-site energy drawn from a uniform 
distribution $[-W/2,W/2]$,
${\hat c}_j^{\dagger}$ is the creation 
operator of an electron at site $j$,
and $t=1$ is the
hopping matrix element.
The repulsive interaction strength is depicted by $U \ge 0$, 
and a positive background is considered.

\section{Non interacting electrons}

\begin{figure}
\onefigure[width=8.5cm,height=!]{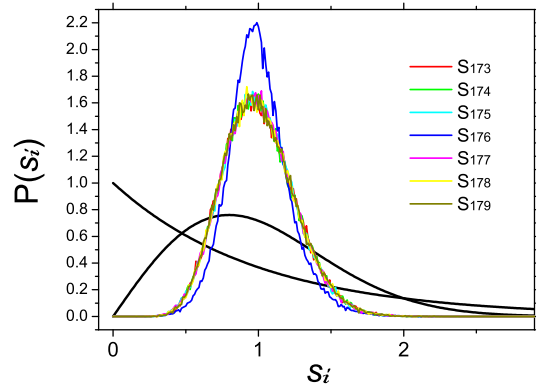}
\caption{The single electron level spacing distribution
$P(s_i)$ for states $i=173 \ldots 179$ around the middle of the
band for a $L=350$, $W=0.3$ wire. For this strength of disorder the
system is disordered 
chaotic $L < \xi$, and the distribution is peaked around the mean
(where the level at the middle of the band $i=176$ is more peaked than
the others).
The Poisson and GOE distributions are provided for comparison.}
\label{fig1}
\end{figure}

All single-electron states of such a 1D system are localized
\cite{lee85}, 
with states at the middle of the band having a localization
length $\xi \approx 105/W^2$
\cite{romer97}. 
We consider a case where the disorder was chosen as
$W=0.3$ and $L=350$, i.e., $\xi \sim 1100 > L$. For this case the single
electron energies $\varepsilon_i$ an eigenstates $\psi_i$ are
readily available via exact diagonalization. Unless otherwise specified,
we perform our analysis over $10,000$ realizations of disorder.
First we consider the single electron level-spacing distribution,
for the $i$-th spacing $\delta_i=\varepsilon_{i+1} -\varepsilon_{i}$.
The unfolded spacing is defined as $s_i= \delta_i / \langle \delta_i \rangle$,
where $\langle \ldots \rangle$ denotes averaging over the different disorder
realization.
The single electron level-spacing will follow the
Poisson statistics ($P_{\rm Poisson}(s_i)=\exp(-s_i)$)
as long as $L \gg \xi$,
while in the metallic (diffusive) regime it should follow
the Wigner (GOE) distribution
($P_{\rm GOE}(s_i)=(\pi s_i/2) exp(-\pi s_i^2/4)$) \cite{shklovskii93}.
For a one-dimensional disordered system there is no genuine metallic
regime since $L < \xi \sim \ell$ (where $\ell$ is the mean free path) 
and therefore the system crosses from a localized to a disordered 
ballistic regime, resulting in a narrower distribution concentrated
around the mean ($s_i=1$). 
This can be clearly seen in Fig. \ref{fig1}, where the
distribution of the single electron level spacings close to the middle
of the band of a ballistic system is presented.

The MBE of the system can not be obtained
by exact diagonaliztion since the size of the many-body
Hilbert space grows as $\left( \begin{array}{c} L \\ N \end{array} \right)$,
which for half-filling ($N=L/2=175$), where $N$ is the number of particles,
is of order $10^{104}$. Thus, another tack is needed. For the non-interacting
case one may calculate the low-lying excitations by considering the 
different permutations of the single electron occupations, $n_i=0,1$, 
of the single electron states. Here, each permutation which obeys
$N=\sum_i n_i$ is a many-body state, with the total many body energy 
$E = \sum_i n_i \varepsilon_i$. These many-body states may be enumerates
by arranging the energies, such that $E_1<E_2<E_3< \ldots$. Of course
there are still an astronomical number of these states and covering them all
is impossible. Nevertheless, if one constrains the search to low lying
excitations, one can limit the the permutations to $p$ (of order
$p \sim E/\delta$) single electron states around the Fermi energy.
Thus, only 
$\left( \begin{array}{c} 2p \\ p \end{array} \right)$ MBE
are considered. Similar to the single particle case the MBE
level spacing  $\Delta_i=E_{i+1} -E_{i}$ can be extracted, as
well as the average $\langle \Delta_i \rangle$ and distribution
$P(S_i)$, where $S_i=\Delta_i/\langle \Delta_i \rangle$.

\begin{figure}
\onefigure[width=8.5cm,height=!]{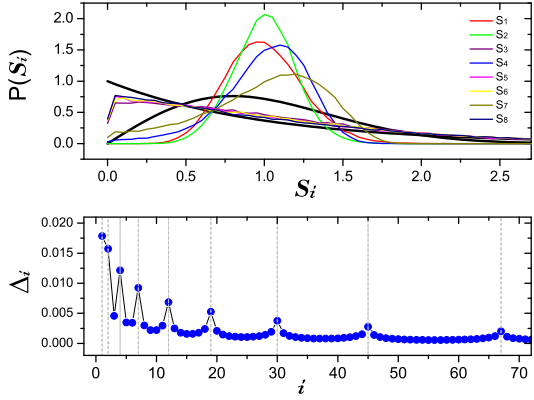}
\caption{The properties of 
the many-body excitation spectrum 
for $10000$ realizations of the non-interacting system  ($U=0$) whose 
single electron level spacings is depicted in Fig. \ref{fig1}. Top panel:
Level spacing distributions for the eight lowest many-particle
levels. Lower panels:
Average spacings as function of the level number. The vertical line depict the
positions of the peaks predicted by the partition function $p(n)$.}
\label{fig2a}
\end{figure}

\begin{figure}
\onefigure[width=8.5cm,height=!]{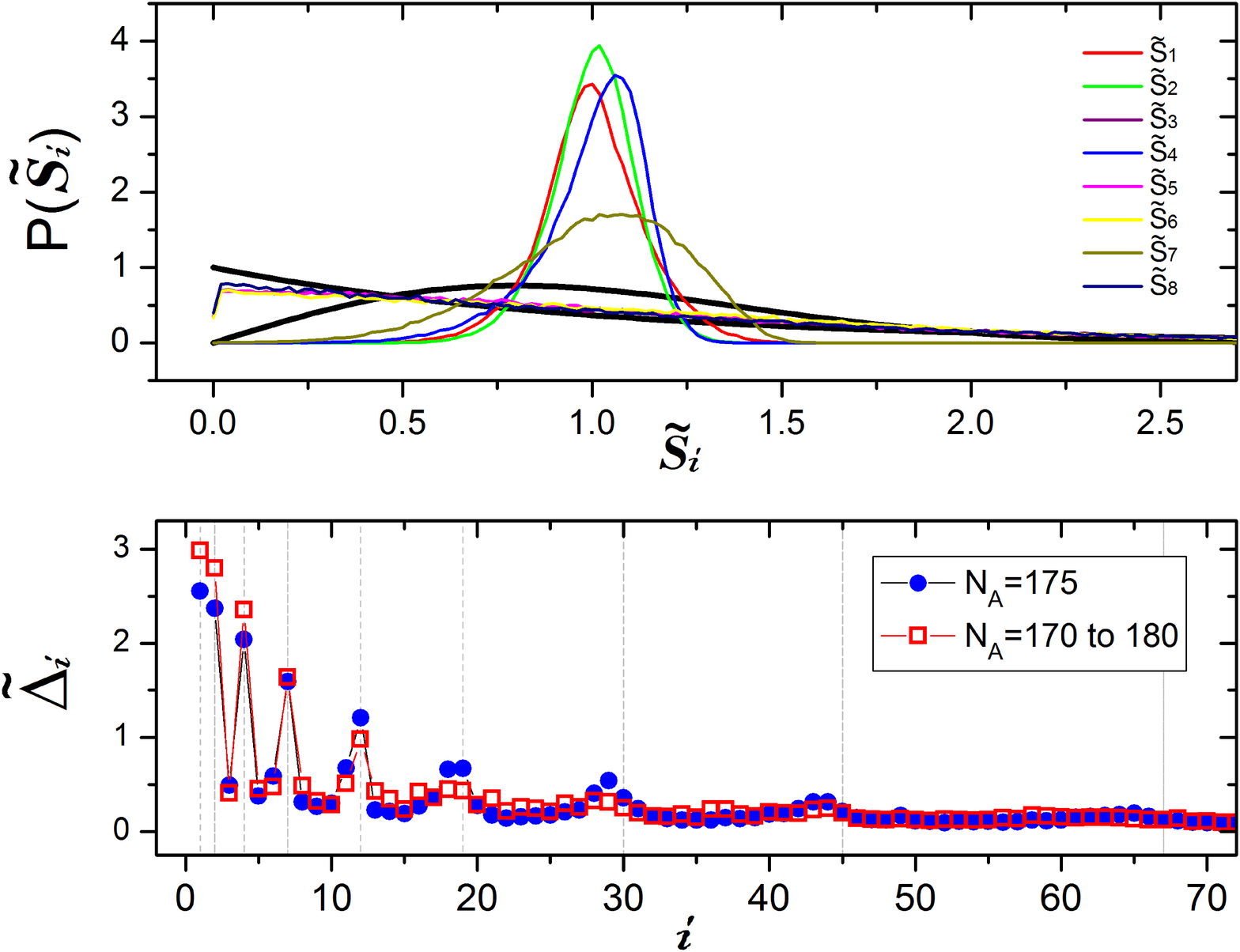}
\caption{The properties of the many body entanglement spectrum 
for an $L_A=350$ region out of a $L=700$ system, with $U=0$, averaged over
$10000$ realizations of disorder.
Top panel:
Entanglement level spacing distributions for the eight lowest levels 
for $N_A=L_A/2$.
Lower panel:
Average spacings as function of the entanglement level number. Blue circles
correspond to $N_A=L_A/2$. Red squares to an additional average over 
$L_A/2-6<N_A<L_A/2+6$.} 
\label{fig2b}
\end{figure}

The results for the average MBE spacings and their 
distribution are plotted on the right panels of Fig. \ref{fig2a}.
It is immediate apparent that the average MBE spacing
exhibits a shell-like (also known as magic numbers in context of
nuclear physics) structure. Spacings $1,2,4,7,12,19,30,45,67 \ldots$ 
are significantly
larger than their neighboring spacings. The reason for this behavior
is combinatorical. The large spacing corresponds to an excitation different
from the previous one by a {\it single} electron moving to the next available
single-electron state.
On the other hand the typical smaller spacings corresponds to the movement
of several electrons between different single-electron states. For equal 
single-level spacings (i.e., $\delta_i=\delta$), the MBE
can acquire energies of $m \delta$ above the ground state 
where $m=1,2,\ldots$. The degeneracy
of the states with energy $m \delta$ depends on the the number of
ways of writing the integer $m$ as a sum of positive integers, 
where the order of addends is not significant.
This corresponds to a problem in combinatorics knows as the
partition problem, and the
answer is given by the partition function $p(m)$
(which is {\it not} the thermodynamical partition function), 
which has a recurrence relation $p(m)=\sum_{k=1}^m
(-1)^{k+1} [p(m-k(3k-1)/2)+p(m-k(3k+1)/2)]$
\cite{andrews98}. 
Thus, $p(m)$ for $m=1,2,\ldots$ are
$1,2,3,5,7,11,15,22,30,42, \ldots$. 
Hardy and Ramanujan have devised an asymptotic form
$p(m) \sim (4 m \sqrt{3})^{-1} \exp(\pi \sqrt{2 m/3})$ \cite{hardy18}.
In the case where the single level
spacing is not constant the degeneracy is lifted, 
but as long as the variation
is not too large, the shell structure
remains and the MBE level spacing between a closed shell and the next
open one is relatively large, as can be seen in Fig. \ref{fig2a}. Thus, 
the number of MBE states between the $j$-th MBE shell (i.e., large level 
spacing) and $j+1$ shell is $p(j)$, in agreement with the numerical
results.

The MBE spacing distributions $P(S_i)$ also retain the shell
structure, where the distributions corresponding to the large spacings are
similar to the single electron spacings (Fig. \ref{fig1}), 
i.e., narrow and peaked around
the mean, while for the other spacings follow the Poisson distribution,
expected when the non-interacting 
MBE differ by more than a single electron
occupation \cite{berkovits94,berkovits96}
(see Fig. \ref{fig2a}).

Does the ES of a finite segment of a disordered system
exhibit similar behavior as might be expected from the Li and Haldanes' 
conjecture
\cite{haldane08}? First we have to calculate the ES,
i.e., the eigenvalues of the RDM. The density matrix 
renormalization group (DMRG) \cite{white92,dmrg}
is a natural candidate for calculating
the ground state of disordered interacting 1D system and
the corresponding eigenvalues of the reduced density matrix. We will use
it for the interacting case. For non-interacting systems, one would expect
that it is possible to extract the eigenvalues without to resort to DMRG.
Direct calculation of the RDM is impossible since its
size is of order $2^{L_A}$ (where $L_A$ is the size of region A), 
so a different approach is needed.

\section{Reduced density matrix eigenvalues using the correlation matrix}

In order to calculate the ESLS for the non-interacting case
we use the connection between the correlation matrix (CM)
and the RDM \cite{Lat2009}. The CM between two sites
in region A is a unitary matrix given by:
\begin{equation}
\label{CMUDa}
C^{\dagger}_{mn}=\left[Tr\{\rho_A c^{\dagger}_{m}c_n\}\right]^\dagger=
Tr\{ c^{\dagger}_{n}c_m\rho_A\}^=C_{nm}.
\end{equation}
One may diagonalize the CM 
and write the trace in terms of the eigenvectors. Thus,
\begin{equation}
\label{CMUDc}
Tr\{\rho_A a^{\dagger}_{q}a_p\}=\nu_q\delta_{qp}
\end{equation}
where $a^\dagger_{q}$ is the 
creation operator of the $q$-th eigenvector and $\nu_q$ is the corresponding 
eigenvalue. 
The RDM for non-interacting electrons 
can be written as an product state of local 
density matrices \cite{Lat2009}:
\begin{equation}
\label{CMUDd}
\rho_A=\rho_1\otimes\rho_2\otimes...\otimes\rho_{L_A}
\end{equation}
where $\rho_q$ is the density matrix associated with the $q$'th 
eigenvalue of the CM.
Moreover, due to the fermionic nature of the particles, $\rho_q$
has to be diagonal. 

The relation between $\nu_q$ and $\rho_q$ is easily extracted \cite{Lat2009}:
\begin{equation}
\label{CMUDe}
\begin{split}A_{qq}=Tr\{\rho_Aa^\dagger_qa_q\}=Tr\{\rho_qa^\dagger_qa_q\}= \\ =Tr\left\{\left(\begin{array}{ccc} 1-\nu_q & 0 \\ 0 & \nu_q \end{array}\right)
\left(\begin{array}{ccc} 0 & 0 \\ 0 & 1 \end{array}\right)\right\}=\nu_q.
\end{split}
\end{equation}
Since $\rho_q$ are independent
any eigenvalue of the
RDM can be constructed by multiplying a permutation of either
$(1-\nu_q)$ (no particle) or $\nu_q$ (one particle) occupying the
$q$-th state. For the case where no particles occupy region A
($N_A=0$), the single eigenvalue of the RDM is
$\lambda^{N_A=0}_1=\prod^{L_A}_{q=1}(1-\nu_q)$.
For the $N_A=1$ the RDM has $L_A$ eigenvalues, where the $i$-th eigenvalue, 
$\lambda^{N_A=1}_i=(1-\nu_1)(1-\nu_2) \ldots \nu_i \ldots (1-\nu_{L_A})$.
Generalizing to any $N_A=p$ particle block in the RDM the first eigenvalue
is constructed by $\lambda^p_1=\nu_1 \ldots \nu_p (1-\nu_{p+1})(1-\nu_{L_A})$,
while the other $\left(\begin{array}{ccc} L_A \\ p \end{array}\right)$
permutations define the rest of the eigenvalues. Since usually many of
the CM eigenvalues are either extremely small or very close to one, 
it is possible to reach an accurate estimation of $\lambda$, with 
significantly fewer permutations.

%
   
The ES obtained for a wire of length $L=700$ and $L_A=350$
occupied by $N_A$ particles and averaged over $10,000$ realizations of disorder.
are presented in Fig. \ref{fig2b}. 
The occupation is $N_A=L_A/2$ or averaged for values of $L_A/2-6<N_A<L_A/2+6$.
This size was chosen in order to correspond to the sizes of the system
for which the MBE were calculated.
For the ES we perform the transformation 
$\epsilon^{N_A}_i=-\ln(\lambda^{N_A}_i)$ and calculate the ESLS 
$\tilde \Delta^{N_A}_i=\epsilon^{N_A}_{i+1} -\epsilon^{N_A}_{i}$, the average 
spacing $\langle \tilde \Delta_i \rangle$ and distribution
$P(\tilde S_i)$, where $\tilde S_i=\tilde \Delta_i/\langle 
\tilde \Delta_i \rangle$. Indeed,
the low-lying ESLS average as well
as the spacing distributions show the same general features 
shown by the MBE. The shell
structure is reproduced for spacings $1,2,4,7,12$,
small deviations in the peak positions appear for the higher spacings at
$19$ and $30$, and is almost wiped up at peaks higher than  
$45$. The general behavior of the ESLS follows the
distribution exhibited by the excitation spacings with narrow 
distributions for spacings $1,2,4$ and $7$, while the 
other spacings follow the Poisson distribution. Nevertheless, details
such as the exact width of the distribution vary between the
excitation and entanglement spacings.

\section{The influence of interactions}

Once interactions are added, our previous arguments do not hold anymore. 
Nevertheless, following the ideas leading to the Fermi liquid picture and 
quasi particles, one expects that for not too strong interactions
and close to the Fermi energy the non-interacting shell picture
will continue to provide a good description of the system. 

\begin{figure}
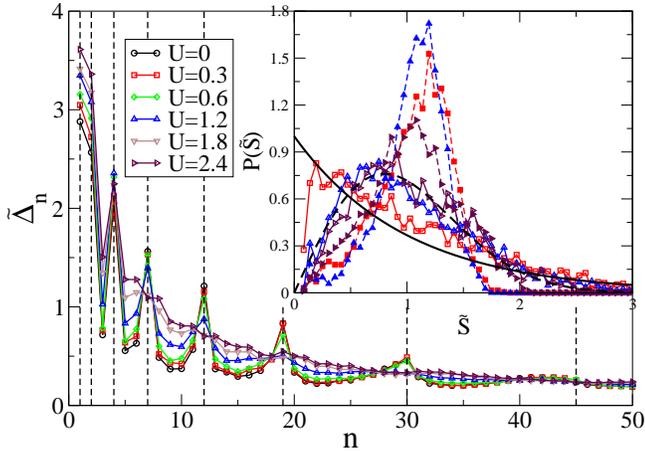

\onefigure[width=8.5cm,height=!]{haldf3.eps}
\caption{Averaged entanglement spectrum level spacings $\tilde \Delta_n$ as
function of the level number $n$ for different interaction strength $U$.
Inset: The spacing, $\tilde S$, distribution for the 3rd (continuous line, 
empty symbols) and
4th (dashed line, full symbols) 
for $U=0.3,1.2$ and $2.4$.
The Poisson (continuous black line) and GOE (dashed black line) are drawn for
comparison.}
\label{fig3}
\end{figure}

As we can not use the combinatoric approach to calculate the 
interacting MBE energy levels, and neither exact diagonaliztion
nor DMRG can provide more than a few excited states, we will
use the correspondence between the MBE and the ES of a finite section 
shown for the non-interacting case to study the evolution of the spectrum as
function of electron-electron interaction strength. Thus we turn on the
interaction $U$ in Eq. (\ref{hamiltonian}) and use DMRG to calculate
the ES over
$100$ realizations of the same disorder, wire length, 
size of region A, and number of particles 
as for the non-interacting case.
The average ESLS  $\tilde \Delta_i$, and distribution
$P(\tilde S_i)$, are calculated and presented in Fig. \ref{fig3}.
For weak interactions $U=0.3$, the shell structure of the average 
spacing is hardly affected.
As interactions increase, the higher shells are washed out, until
for $U=2.4$ only spacings $1,2$ and $4$ remain larger than their neighbors.
The entanglement spacing distribution is also transformed as function 
of interaction strength. For weak interactions ($U=0.3$)
the distribution $P(\tilde S_3)$ is close to Poisson, while the distribution
for the $4$-th  (large) spacings, $P(\tilde S_4)$, is peaked around the mean,
as for the non-interacting case (see inset Fig. \ref{fig3}). 
This distinction is blurred as the interactions become stronger and
$P(\tilde S_i)$ for any $i$ approaches the GOE distribution.  
This is in agreement with the observations for the distribution
of interacting MBE
\cite{berkovits94,berkovits96,pascaud98,berkovits99,song00,oganesyan07}
which show a transition to GOE statistics as interactions become stronger.
Thus, it seems that the Li and Haldane conjecture holds even for the ES 
of interacting systems.

\section{Discussion}

Thus, Li and Haldanes' conjecture on the correspondence between the 
low-lying MBE spectrum, and its
ES is confirmed for non-interacting disordered systems. Turning on
interactions gradually change the ES of the system in a manner that fits
the expectations for the behavior of the MBE. We conclude that ES is
a promising route to study MBE properties beyond the first few excitations, 
which are impossible to
study by any other way for large systems. This could be a promising
way to study phenomena occurring only for the MBE, such as the many-body
localization transition \cite{basko06}. Although still limited to low-lying
excitations, the fact that higher many-body excitation may be probed improves
the possibility to glean useful information on the transition.
Indeed in the behavior of the shell structure
as function of interaction strength presented in Fig. \ref{fig3}, where
as interaction increases lower shell peaks disappear, 
may show a glimpse of this phenomena.

For non-interacting systems we have presented a computational effective
method to extract the
low-lying ES from the CM. This method could be used for the study of the
ES for higher dimension, which is otherwise quite daunting.

We have also shown a new shell (magic number) structure appearing for 
disordered ballistic
one dimensional systems and explained its origin. The low-lying shells
are robust and are not wiped out by interactions. It is interesting to
understand whether this behavior survives for stronger disorder where the
localization length is smaller than the length of region A. On one hand, since
the disorder is larger it is expected to wipe out the shell structure, on the
other hand, the area in region A entangled with the rest of the system
remains of order of $\xi$ and thus shrinks with the disorder, i.e., one 
effectively samples a smaller sample. Also the study of higher dimensions
may prove interesting.

\acknowledgments

We would like to thank B. L. Altshuler for useful discussions and
the Institute of Basic Science Center for Theoretical Physics of 
Complex Systems, Daejeon, South Korea where part of this work was performed.




\end{document}